\documentclass[aps,pra,twocolumn,superscriptaddress,amsmath,showpacs,amssymb]{revtex4}
\usepackage{bm}
\usepackage{amsmath,amssymb}
\usepackage[dvips,final]{graphicx}

\newcommand{\be}{\begin{equation}}
\newcommand{\ee}{\end{equation}}
\newcommand{\BM}{\begin{pmatrix}}
\newcommand{\EM}{\end{pmatrix}}

\renewcommand{\phi}{\varphi}
\newcommand{\MH}{\mathrm{H}}
\newcommand{\bt}{\Tilde{b}}
\newcommand{\bb}{\Bar{b}}
\newcommand{\xit}{\Tilde{\xi}}
\newcommand{\xib}{\Bar{\xi}}
\newcommand{\phib}{\Bar{\phi}}
\renewcommand{\d}{\dagger}
\newcommand{\ve}{\varepsilon}
\newcommand{\bra}{\langle 0 |}
\newcommand{\ket}{| 0\rangle}
\newcommand{\Real}{\mathrm{Re}}
\newcommand{\Imag}{\mathrm{Im}}
\newcommand{\INT}{\int\!\!}
\newcommand{\bdgL}{\mathcal{L}}
\newcommand{\bdgM}{\mathcal{M}}
\newcommand{\tpartial}{\frac{\partial}{\partial t}}
\bmdefine{\bx}{x}
\bmdefine{\by}{y}
\bmdefine{\bz}{z}
\bmdefine{\bchi}{\chi}
\bmdefine{\bY}{Y}
\begin{document}

\title{Unifying treatment of nonequilibrium and unstable dynamics of cold bosonic atom system with time-dependent order parameter 
in Thermo Filed Dynamics}
%
\author{Y.~Nakamura}
\email{nakamura@aoni.waseda.jp}
\affiliation{Department of Electronic and Photonic Systems, Waseda
University, Tokyo 169-8555, Japan} 
\author{Y.~Yamanaka}
\email{yamanaka@waseda.jp}
\affiliation{Department of Electronic and Photonic Systems, Waseda
University, Tokyo 169-8555, Japan} 

\begin{abstract}
The coupled equations which describe the temporal evolution of the Bose-Einstein condensed system 
are derived in the framework of nonequilibrium Thermo Field Dynamics.
The key element is that they are not the naive assemblages of presumable equations, 
but are the self-consistent ones derived by appropriate renormalization conditions.
While the order parameter is time-dependent, an explicit quasiparticle picture 
is constructed by a time-dependent expansion. 
Our formulation is valid even for the system with a unstable condensate, 
and describes the condensate decay caused by the Landau instability as well as by the dynamical one.
\end{abstract}


\maketitle

\section{Introduction}
The systems of trapped cold atoms are ideal for studying the foundations of quantum
many-body theories such as quantum field theory and thermal field
theory. They are dilute and weak-interacting, so theoretical
calculations can be compared with experimental results directly. 
Since the realization of Bose--Einstein condensates \cite{Cornell,Ketterle,Bradley}, 
many intriguing phenomena have been observed with good accuracy,
and offer opportunities to test many aspects of quantum many-body theories in both
equilibrium and nonequilibrium. 
Among them, the unstable phenomena of the condensate attract our attention, 
because firstly to formulate unstable quantum many-body systems is still an open problem and
and secondly nonequilibrium processes accompany the instability in thermal situation.

Theoretically, the instability of the condensate is characterized by 
the eigenvalue of the Bogoliubov-de Gennes (BdG) equations \cite{Bogoliubov, deGennes, Fetter}, 
which follow from linearization of the time-dependent Gross-Pitaevskii (TDGP) equation \cite{Dalfovo}.
Since the BdG equations are generally eigenvalue ones for non-Hermitian matrices 
, their eigenvalues can be complex. 
The emergence of complex eigenvalues is interpreted as the sign of the dynamical instability.
This instability is associated with the decay of the initial configuration of the condensate 
and can occur even at zero temperature.
On the other hand, if the negative eigenvalues for a positive-norm mode are present, 
the system shows another instability, called the Landau instability.
This instability, in which the thermal cloud plays an essential role to 
drive the condensate toward a lower energy state, is suppressed at very low temperature.

The observations of both the Landau and dynamical instabilities are reported in several systems, 
especially in the system where the condensate flows in an optical lattice \cite{Fallani1, Fallani2},
and they are in good agreement with the analyses of the BdG equations \cite{Wu1, Wu2}.

Although the TDGP equation outlines the experiments corresponding to the dynamical
instability at very low temperature \cite{Huhtamaki, Munoz},
 e.g. the multiply-quantized vortex splitting \cite{Shin}, 
an detailed description of the unstable dynamics in thermal situations is not trivial.
That is because there is no more quasi-stable state, and so a fully nonequilibrium theory is required.

There are known two nonequilibrium thermal field theories,
i.e., the closed time path (CTP) formalism \cite{CTP} 
and the Thermo Field Dynamics (TFD) \cite{AIP}. 
The CTP formalism is widely used. But we employ the TFD
formalism in this paper, because the concept of quasiparticle picture 
which is essential for quantum field theory is clear even in nonequilibrium situations.
In TFD, which is a real-time canonical formalism of quantum field
theory, thermal fluctuation is introduced through doubling the degrees
of freedom, and the mixed state expectation in the density matrix formalism 
is replaced by an average of a pure state vacuum, called the thermal vacuum. 
It is crucial in our formulation of TFD to construct the interaction picture. 
In quantum field theory, the choice of unperturbed Hamiltonian and fields 
is that of quasiparticle  picture, and concrete calculations are possible 
only when a particular unperturbed representation, or a particular
particle picture, is specified. One does not know an exact unperturbed
representation beforehand. 

So far we have investigated the cold atom system with a time-independent 
configuration of  the condensate in TFD, and derived the quantum transport equation which 
describes the temporal evolution of the quasiparticle number distribution \cite{NSMOY}.
It was essential to construct an explicit quasiparticle picture there.
In contrast to the previous investigations \cite{Stoof,Zaremba,Gardiner2,Miesner,Yamashita} 
which are based on a phase-space
distribution function, our transport equation contains an additional collision term which 
is traced back to our choice of an appropriate quasiparticle picture. 
The additional collision term, which we call the triple production term, vanishes in the equilibrium limit 
if there is no Landau instability, but remains non-vanishing to prevent the system from 
equilibrating if there is Landau instability.
Thus our transport equation with the triple production term
and the other ones without it predict definitely different behaviors of the unstable system.

In this paper, we derive the coupled equations which describe 
the nonequilibrium dynamics of the cold atom system with a time-dependent order parameter. 
They are the TDGP equation, the TDBdG equations, and the quantum transport equation. 
The key points are that while the order parameter is time-dependent,
we construct a time-independent quasiparticle picture and so that the stable vacuum 
which are essential for quantum field theory.
These are accomplished by expanding the field operator with the time-dependent complete set 
evaluating by the TDBdG equations \cite{Matsumoto1}.
The quantum correction to the TDGP equation is determined 
self-consistently and simultaneously as the quantum transport equation by some renormalization conditions \cite{Chu3}.
Solving the coupled equations numerically, we illustrate the dynamics of the condensate decays with either the 
Landau instability or the dynamical one and discriminate the two instabilities.

This paper is organized as follows. 
We consider the cold bosonic atom system with a time-dependent order parameter at zero temperature in Section II.
We show that it is crucial to expand the field operator by the solutions of TDBdG equations 
to maintain the time-dependent quasiparticle picture.
In Section III, a nonequilibrium system is considered, 
and the degrees of freedom are doubled to treat the system with the nonequilibrium TFD.
We construct a systematical method to obtain the coupled equations, 
and derive those explicitly in the leading order.
In Section IV, we consider a simple system with the Bose--Hubbard model and calculated the coupled equations numerically.
Section V is devoted to summary.

\section{Formulation of quantum field theory}
In this section, we consider the cold bosonic atom system with a time-dependent order parameter at zero temperature.
We start with the following Hamiltonian to describe the trapped dilute bosonic atoms:
\begin{multline}
	H = \INT d^3x \biggl[ \psi^\d(x)\left( -\frac{\nabla^2}{2m} + V(\bx) -\mu \right)\psi(x) \\
	+ \frac{g}{2}\psi^\d(x)\psi^\d(x)\psi(x)\psi(x)\biggr] \,,
\end{multline}
where $m$, $V(\bx)$, $\mu$, and $g$ represent the mass of an atom, the trap potential, 
the chemical potential, and the coupling constant, respectively. 
Throughout this paper $\hbar$ is set to be unity.
The bosonic field operator $\psi(x)$ obeys the canonical commutation relations 
\begin{align}
	\left[ \psi(x), \psi^\d(x') \right] |_{t=t'} &= \delta(\bx-\bx') \,,\\
	\left[ \psi(x), \psi(x') \right] |_{t=t'} &= 0 \,.\,
\end{align}
where $x = (\bx, t)$.
Reflecting the existence of the condensate, the field operator $\psi$ is divided 
into a classical part $\zeta(x)$ and a quantum one $\phi(x)$ on the criterion $\bra\phi(x)\ket=0$. 
Note that the vacuum is not yet specified and that $\zeta(x)$ is an arbitrary
function at this stage, 
and the division must be completed later self-consistently.
The doublet notation is introduced as
\be
	\phi^\alpha = \BM \phi \\ \phi^\d \EM^\alpha \,,\qquad
	\phib^\alpha = \BM \phi^\d & - \phi \EM^\alpha\,,
\ee
and the unperturbed Hamiltonian $H_0$, bilinear and linear in $\phi$ and $\phi^\d$, is 
\begin{multline} \label{BEC_H_0}
	H_0 = \INT d^3x\, \Bigl[\, \frac{1}{2} \phib^\alpha T^{\alpha\beta} \phi^\beta 
	+  \phi^\d\left( h_0\zeta  +\delta C\right)\\
	+ \phi\left(h_0 \zeta^* +\delta C^* \right)\Bigr]  \,,
\end{multline}
with
\be
	T(x) = T_0(x) + \delta T(x)\,,
\ee
where
\begin{align}
	T^{\alpha\beta}_0(x) &= \BM \bdgL_0(x) & \bdgM_0(x) \\ -\bdgM_0^*(x) & -\bdgL_0(x)
	 \EM^{\alpha\beta} \,,\\
	\bdgL_0(x) &= -\frac{\nabla^2}{2m} + V(\bx) -\mu + 2g|\zeta(x)|^2 \,,\\
	\bdgM_0(x) &= g\zeta^2(x) \, ,\\
	h_0(x) &= -\frac{\nabla^2}{2m} + V(\bx) -\mu + g|\zeta(x)|^2\,.
\label{eq:defh0}
\end{align}
The counter terms $\delta T(x)$ and $\delta C(x)$ are determined later self-consistently.
The perturbed Hamiltonian $H_{\mathrm{int}} = H-H_0$ is given as
\begin{multline}
	H_{\mathrm{int}} = \INT d^3x \biggl[\frac{g}{2}\phi^{\d,2}\phi^2 + g\zeta^*\phi^\d\phi^2 \\
	+ g\zeta\phi^{\d,2}\phi - \phi^\d \delta C - \phi \delta C^*\biggr] \,.
\end{multline}

From the original Heisenberg equation for $\psi$ and the time-dependent $\zeta$,
the field equation for $\phi$ in the interaction picture should be
\be \label{HeigenbergEqforPhi}
	i\Dot{\phi} = \left( \bdgL_0 +\delta T^{11}\right)\phi 
	+ \left(\bdgM_0+  \delta T^{12}\right) \phi^\d + h_0 \zeta +\delta C -i\Dot{\zeta} \,.
\ee
Due to the last term, this time-evolution is generated not by $H_0$
in Eq.~(\ref{BEC_H_0}) but by
\be
	H_0^\phi = H_0 - i\INT d^3x \left[ \Dot\zeta\phi^\d - \Dot\zeta^*\phi \right] \,,
\ee
as
\be
	i\Dot{\phi}= \left[ \phi, H_0^\phi \right] \, .
\ee
The condition $\bra\phi(x) \ket=0$ must hold for any $t$ in the unpertubed representation,
which implies 
\be
i\frac {\partial}{\partial t}\bra\phi(x) \ket=\bra i{\dot \phi} \ket=0 \,. 
\ee
for the time-independent vacuum. According to Eq.~(\ref{HeigenbergEqforPhi}), we have
\be \label{eq:deltaC}
\delta C = (i\partial_t -h_0)\zeta \,,
\ee
and the time-evolution operator $H_0^\phi$ becomes a simple quadratic form:
\be
	H_0^\phi = \frac{1}{2} \INT d^3x \; \phib^\alpha T^{\alpha\beta} \phi^\beta \,.
\ee

\subsection{Field expansion for time-independent order parameter}
Before going into further discussions, let us briefly review the Bogoliubov-de Gennes (BdG) method 
which diagonalizes the unperturbed Hamiltonian $H_0$ ($=H_0^\phi$) in case of 
the time-independent order parameter $\zeta(\bx)$.
The BdG equations are simultaneous eigenvalue equations given by \cite{Bogoliubov,deGennes,Fetter}
\be 
	T(\bx) \by_\ell(\bx)  =  \omega_\ell \by_\ell(\bx) \,.
\ee
Since the operator $T$ is non-Hermitian, the eigenvalues is not always real but can be complex
in general.
The condition for the emergence of complex eigenvalues 
in the BdG equations has been studied both
numerically \cite{Pu,Wu,Kawaguchi} and analytically \cite{Skryabin,Taylor,Nakamura}, 
and the quantum field theoretical formulation has also been discussed \cite{Mine}. 
The emergence of complex eigenvalues implies the dynamical instability of the system, 
and a drastic temporal change of the order parameter occurs then.
While our propose in this subsection is to find a stable initial condition
 as becomes apparent later, 
we consider only the case where no complex eigenvalue emerges. 

Eigenfunctions belonging to the non-zero real eigenvalues can be
orthonormalized under the indefinite metric as
\begin{align} 
	\INT d^3x\; \by_{\ell}^\d(\bx) \sigma_3 \by_{\ell'}(\bx) &=
 \delta_{\ell\ell'} \,,\label{orthonormal_1}\\ 
	\INT d^3x\; \bz_{\ell}^\d(\bx) \sigma_3 \bz_{\ell'}(\bx) &=
 -\delta_{\ell\ell'} \,,\\ 
	\INT d^3x\; \by_{\ell}^\d(\bx) \sigma_3 \bz_{\ell'}(\bx) &=
 0 \,,\label{orthonormal_3} 
\end{align}
with $i$-th Pauli matrix $\sigma_i$. The function $\bz_\ell$, defined by
$\bz_\ell = \sigma_1 \by_\ell^*$, is an eigenfunction belonging to
$-\omega_\ell$, when $\by_\ell$ is an eigenfunction belonging to $\omega_\ell$. 

Due to the Nanbu-Goldstone theorem \cite{NGtheorem}, 
there is  a zero mode eigenfunction in the BdG equations \cite{Matsumoto2, Lewenstein}:
$
	T \by_0 = 0 \,.
$
The zero mode eigenfunction $\by_0$ is orthogonal to all the other eigenfunctions,
and what is more, is orthogonal to itself.
Hence, an additional adjoint mode $\by_{-1}$ has to be introduced for the completeness as
\be
	T \by_{-1} = I \by_0 \,,
\ee
where $I$ is determined to satisfy the normalization condition: 
\be
	\INT d^3x\; \by_{-1}^\d(\bx) \sigma_3 \by_{0}(\bx) = 1 \,.
\ee

It is convenient to rewrite the whole orthonormal conditions with the $2\times2$ matrix
form as 
\be	\label{eq:Wortho}
	\INT d^3x\; W_{\Lambda}(\bx) \; W_{\Lambda'}^{-1}(\bx) = \delta_{\Lambda\Lambda'}\,, 
\ee
where
\begin{align} \label{eq:defW}
	W_\ell(\bx) &= \sigma_3 \BM \by_\ell^\d(\bx) \\ \bz_{\ell}^\d(\bx) \EM \sigma_3 \,,&
	W_\ell^{-1}(\bx) &= \BM \by_\ell(\bx) & \bz_\ell(\bx) \EM \,,\\
	W_0(\bx) &= \sigma_1 \BM \by_0^\d(\bx) \\ \by_{-1}^\d(\bx) \EM \sigma_3 \,,&
	W_0^{-1}(\bx) &= \BM \by_0(\bx) & \by_{-1}(\bx) \EM \,,
\end{align}
with $\Lambda = \ell,\,0.$ 

The completeness condition, 
\begin{align}
	&\hspace{1.5cm}\sum_{\ell} \left[\by_\ell(\bx) \by_\ell^\d(\bx')- \bz_\ell(\bx) \bz_\ell^\d(\bx')\right] \notag\\
	&+ \by_0(\bx) \by_{-1}^\d(\bx')+ \by_{-1}(\bx) \by_{0}^\d(\bx') 
	=\sigma_3 \delta(\bx-\bx')\,,
\end{align}
is simply expressed as
\be \label{eq:Wcompl}
	\sum_\ell W_\Lambda^{-1}(\bx) \; W_\Lambda(\bx') = \delta(\bx-\bx') \,,
\ee
and the field operators in the doublet form are expanded   as
\begin{align}
	\phi^\alpha(x) &= \sum_\Lambda W_\Lambda^{-1, \alpha\beta}(\bx)
 b_\Lambda^\beta(t) \,,\\
	\Bar\phi^\beta(x) &= \sum_\Lambda \bb_\Lambda^\alpha(t)
 W_\ell^{\alpha\beta}(\bx) \,, 
\end{align}
where
\begin{align}
	b_\ell^\alpha &= \BM b_\ell \\ b^\d_\ell \EM^\alpha \,,&
	\bb_\ell^\alpha &= \BM b^\d_\ell & -b_\ell\EM^\alpha \,,\\
	b_0^\alpha &= \BM -iq \\ p \EM^\alpha \,,&
	\bb_\ell^\alpha &= \BM p & iq \EM^\alpha \,.
\end{align}
The operators $b_\ell$, $p$, and $q$ satisfy the canonical commutation relations 
$[ b_\ell , b_{\ell'}^\d ] = \delta_{\ell\ell'}$ and $[q, p] =i$, respectively,  
and the unperturbed Hamiltonian becomes
\begin{align}
	H_0^\phi &= H_0 = \frac{1}{2} \int\!\! d^3x\; \Bar\phi^\alpha(x)
 \;T^{\alpha\beta}({\bx}) \; \phi^\beta(x) \\
		&= \sum_\ell \omega_\ell b_\ell^\d b_\ell + \frac{p^2}{2}\,,
\end{align}
which is diagonalized except for the zero mode sector, that is, that of the
 quantum mechanical operators $p$ and $q$. The choice of the wave function for the
sector is not settled yet. 
Although the diagrammatic calculation is possible once a particular wave function for the sector
 is chosen,  we suppress the zero mode in what follows in order to avoid the ambiguity.

\subsection{Field expansion for time-dependent order parameter}

The time-dependent order parameter $\zeta(x)$ implies the time-dependent
$T(x)$.  In order to deal with this situation, we suppose a 
time-dependent orthonormal complete set $\{W_\ell(x)\}$ which is defined
by Eq.~(\ref{eq:defW}) with some time-dependent functions $\by_\ell(x)$ and $\bz_\ell(x)$
and which has the properties of Eqs.~(\ref{eq:Wortho}) and (\ref{eq:Wcompl}) at equal time.
Then the field operators are expanded as
\be \label{expand_phi2b}
	\phi^\alpha(x) = \sum_\ell W_\ell^{-1, \alpha\beta}(x) b_\ell^\beta(t) \,.
\ee

We note that because of the time-dependence of $W_\ell(x)$ the time-evolution operator for $b_\ell$ is not $H_0^\phi$  but $H_0^b$:
\begin{align}
	&H_0^b(t) = H_0^\phi(t)  \notag\\
	&\hspace{0.5cm}
	-\frac{i}{2}\INT d^3x d^3x'\; \phib^\alpha(x) 
	\biggl[ \sum_\ell \Dot{W}^{-1}_\ell(x) W_\ell(x')\biggr]^{\alpha\beta} \hspace{-0.2cm}\phi^\beta(x') 
\end{align}
Here and hereafter, we take a common time variable for $x=({\bx},t)$ and $x'=(\bx', t)$.
As we are going to treat the time-dependent order parameter in the quasi-particle picture represented by $b_\ell$, the necessary condition is that $H_0^b$ to be diagonal:
\begin{align}
	H_0^b(t) &= \sum_\ell \lambda_\ell(t) b^\d_\ell(t) b_\ell(t) \\
	&= \frac{1}{2} \sum_\ell \bb_\ell^\alpha(t) \left[ \lambda_\ell(t) \sigma_3 \right]^{\alpha\beta}b_\ell^\beta(t) \,,
\end{align}
where $\lambda_\ell(t)$ is an arbitrary real function.
Therefore, $W_\ell(x)$ must satisfy
\be
	T W^{-1}_\ell = i\tpartial{W}^{-1}_\ell + \lambda_\ell W^{-1}_\ell \sigma_3 \,.
\ee
We eliminate $\lambda_\ell$ from the equations by the replacement 
$W_\ell^{-1}(x) \to W_\ell^{-1}(x)\, e^{i\INT^t ds\; \lambda_\ell(s)\sigma_3}$, and obtain
\be \label{TDBdGforW}
	T(x) W^{-1}_\ell(x) = i\tpartial{W}^{-1}_\ell(x) \,,
\ee
or equivalently the TDBdG equations
\be	\label{TDBdGfory}
	T(x) {\by}_\ell(x) = i\tpartial{\by}_\ell(x) \,.
\ee
If the orthonormal complete set $\{W_\ell(x)\}$ is chosen as the initial condition 
of Eq.~(\ref{TDBdGforW}), 
it keeps the orthonormality and the completeness for all the 
time because of the following equations:
\begin{align}
	i\frac{d}{dt} \INT d^3x\; W_\ell(x) W_{\ell'}^{-1}(x) &= 0 \,,\\
	i\tpartial \sum_\ell W_\ell^{-1}(x) W_\ell(x') &= 0\,.
\end{align}

Thus the procedure of constructing the time-dependent complete set is obtained: 
Solve the eigenvalue problem $T(x)W_\ell(x) = \omega_\ell W_\ell(x)$ at an initial time $t_0$, 
and calculate the time evolution according to $i\tpartial W_\ell(x) = T(x)W_\ell(x)$.
Next, expand the field operator $\phi(x)$ by the complete set $\{W_\ell(x)\}$ 
as Eq.~(\ref{expand_phi2b}). This expansion is obviously reduced to the ordinary
one in the limit of time-independent order parameter and has already been proposed
by Matsumoto and Sakamoto \cite{Matsumoto1}.
What we have shown in the above paragraph is that the choice of $W_\ell(x)$ is 
justified from the viewpoint of quantum field theory: the quasi-particle operator $b_\ell$,
constructing the Fock space on the the Bose--Einstein condensed vacuum, diagonalizes
the time evolution operator even for the time-dependent order parameter.

\section{Nonequilibrium TFD formulation}
In this section, we double every degree of freedom 
to treat a nonequilibrium system in  TFD and introduce the thermal Bogoliubov transformation
\be
	\BM b_\ell \\ \bt_\ell^\d \EM = B_\ell^{-1} \BM \xi_\ell \\
 \xit_\ell^\d \EM  \,,\qquad
	\BM b^\d_\ell & -\bt_\ell \EM = \BM \xi^\d_\ell & -\xit_\ell
 \EM B_\ell \,,
\ee
with
\begin{align}
	B_\ell^{\mu\nu}(t) &= \BM 1+ n_\ell(t) & - n_\ell(t) \\ -1
                                & 1 \EM^{\mu\nu} \,,\\
	B^{-1, \mu\nu}_\ell(t) &= \BM 1 &  n_\ell(t) \\ 1 & 1+ n_\ell(t)
        \EM^{\mu\nu} \,. 
\end{align}
It is important to take the above particular form of the thermal
Bogoliubov matrix, as one calls the $\alpha=1$ representation \cite{AIP},
which enables us to make use of the Feynman diagram method in
nonequilibrium systems \cite{Evans}. 
In TFD, the thermal average is represented by the pure state expectation of the 
thermal vacuum, denoted by $\ket$, and the operators who annihilate 
$\ket$, are not the $b$-operators but the $\xi$-ones:
\be
	 \xi_\ell\ket = \xit_\ell \ket = 0\,,\qquad
	 \bra \xi_\ell^\d = \bra \xit_\ell^\d = 0\,.
\ee
The number distribution $n_\ell(t)$ is given by 
\be
	n_\ell(t) = \bra b_\ell^\d(t) b_\ell(t) \ket \,,
\ee 
and its time dependence is determined later.
The combination of the two transformations, $\xi$ into $b$ and $b$ into $\phi$, 
involves the $4\times4$-matrix transformations, 
\begin{align}
	b_\ell^{\mu\alpha} &= \mathcal{B}_\ell^{-1, \mu\alpha\nu\beta}
 \xi_\ell^{\nu\beta} \,,& 
	\bb_\ell^{\nu\beta} &= \xi_\ell^{\mu\alpha}
 \mathcal{B}_\ell^{\mu\alpha\nu\beta} \,,\\ 
	\phi^{\mu\alpha} &= \sum_\ell \mathcal{W}_\ell^{-1,
 \mu\alpha\nu\beta} b_\ell^{\nu\beta} \,,& 
	\Bar\phi^{\nu\beta} &= \sum_\ell \bb_\ell^{\mu\alpha}
 \mathcal{W}_\ell^{\mu\alpha\nu\beta} \,, 
\end{align}
with the $4\times4$ thermal Bogoliubov and BdG matrices
\begin{align}
	\mathcal{B}_\ell^{-1, \mu\alpha\nu\beta} 
	&= \delta_{\alpha 1} \delta_{\beta 1} B_\ell^{-1, \mu\nu}
	+ \delta_{\alpha 2} \delta_{\beta 2}
	\left(\sigma_1 B_\ell^{-1}\sigma_1\right)^{\mu\nu} \!,\\
	\mathcal{B}_\ell^{\mu\alpha\nu\beta} &= \delta_{\alpha
 1}\delta_{\beta 1} B_\ell^{\mu\nu}  
	+ \delta_{\alpha 2}\delta_{\beta 2}\left(\sigma_1
 B_\ell\sigma_1\right)^{\mu\nu}  \,,\\ 
	\mathcal{W}_\ell^{-1, \mu\alpha\nu\beta} &= \delta_{\mu \nu}
 W_\ell^{-1,\alpha\beta} \,,\\ 
	\mathcal{W}_\ell^{\mu\alpha\nu\beta} &= \delta_{\mu
 \nu}W_\ell^{\alpha\beta}\,,
\end{align}
where the quartet notations for $b_\ell$ are introduced by
\begin{align}
	b_\ell^{\mu\alpha} 
	&= \BM b_\ell^\mu \\[8pt] [\sigma_1\bt_\ell]^\mu \EM^\alpha
	= \BM b_\ell \\ \bt_\ell^\d \\ b_\ell^\d \\ \bt_\ell
 \EM^{\mu\alpha} \,,\\ 
	\bb_\ell^{\nu\beta}
	&= \BM \bb_\ell^\nu & [\Tilde\bb_\ell \sigma_1]^\nu \EM^\beta 
	= \BM b_\ell^\d & -\bt_\ell & -b_\ell & \bt_\ell^\d
 \EM^{\nu\beta} \,, 
\end{align}
and in similar fashions for $\xi_\ell$ and $\phi$.
The Hamiltonian of TFD, which should generate the time translations of both the 
non-tilde and tilde operators, 
is not the ordinary Hamiltonian $H$ but the hat Hamiltonian $\Hat{H} = H - \Tilde{H}$.
The time-independence of the thermal vacua requires the minus sign in front of $\Tilde{H}$.
Furthermore, the unperturbed Hamiltonian in nonequilibrium TFD is 
$\Hat{H}_Q = \Hat{H}_0-\Hat{Q}$ for nonequilibrium system with 
$\Hat{H}_0 = H_0 - \Tilde{H}_0$ 
and the thermal counter term $\Hat{Q}(t)\,$, caused by the time-dependence of $n_\ell(t)$:
\be
	\Hat{Q}(t) = -\frac{i}{2}\sum_\ell \Dot{n}_\ell(t) \, 
	\xib_\ell^{\mu\alpha}(t) 	\left(\begin{array}{cc|cc}
		 0&1& & \\ 0&0 & & \\ \hline & & 0 & 0 \\ & &1 &0 
	\end{array}\right)^{\mu\alpha\nu\beta} \xi_\ell^{\nu\beta}(t) \,.
\ee

The unperturbed and full propagators for $\phi$ and $\xi$ are defined by
\begin{align}
	\Delta^{\mu\alpha\nu\beta}(x_1, x_2) &= -i \bra
 T[\phi^{\mu\alpha}(x_1) \Bar\phi^{\nu\beta}(x_2)] \ket \,,\\ 
	G^{\mu\alpha\nu\beta}(x_1, x_2) &= -i \bra
 T[\phi_\MH^{\mu\alpha}(x_1) \Bar\phi_\MH^{\nu\beta}(x_2)] \ket \,,\\ 
	d^{\mu\alpha\nu\beta}_{\ell\ell'}(t_1,t_2) &= -i\bra
 T[\xi^{\mu\alpha}_{\ell}(t_1) \xib^{\nu\beta}_{\ell'}(t_2)] \ket \,,\\ 
	g^{ \mu\alpha\nu\beta}_{\ell\ell'}(t_1,t_2) &= -i\bra
 T[\xi^{\mu\alpha}_{\MH\ell}(t_1) \xib^{\nu\beta}_{\MH\ell'}(t_2)] \ket
 \,,
\end{align}
respectively, where the subscript $\MH$ denotes a quantity in the Heisenberg picture.
The self-energies $\Sigma$ and $S$ are introduced by the Dyson equations 
$G = \Delta + \Delta\Sigma G$ and $g = d + dSg$, respectively.

Our critical step is to adopt the following three renormalization conditions simultaneously
to determine the whole time evolution of the system, explicitly to determine the unknown functions
$n_\ell(t)$, $\delta T(x)$ and  $\zeta(x)$ ( or $\delta C(x)$, see Eq.~(\ref{eq:deltaC})):
\begin{alignat*}{2}
	&\text{(i)}\quad & g^{1121}_{\ell\ell}(t,t) &= 0\,\\
	&\text{(ii)}\quad & \Real\, S^{1111}_{\ell\ell, \mathrm{on-shell}} &= 0\,,\\
	&\text{(iii)}\quad & \bra \phi_\MH(x) \ket &= 0\,.
\end{alignat*}

\begin{figure}
\includegraphics[width=6cm]{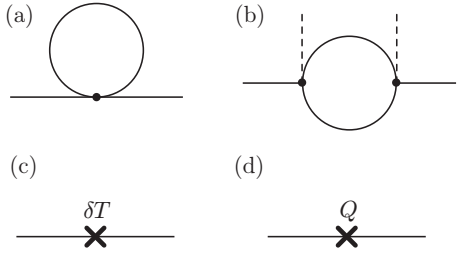}
\caption{\footnotesize{%
One-loop propagators and counter terms.
The solid and dashed lines denote the unperturbed propagator and the order parameter $\zeta$,
 respectively.
}}
\label{fig:propagator}
\end{figure}

The condition (i) is what we have proposed for a nonequilibrium system with the static
 condensate \cite{NSMOY} as a natural extension of the one for non-condensed system
 proposed by Chu and Umezawa \cite{Chu1, Chu2, Chu3}.  
It provides the transport equation which determines the temporal evolution 
of the unperturbed number distribution $n_\ell(t)$.  The possible 
diagrams in the leading order are indicated in Fig.~\ref{fig:propagator}.
Because the contributions from Fig.~\ref{fig:propagator} (a) and (c) vanish, 
the leading ones come from Fig.~\ref{fig:propagator} (b) and (d), and the latter one is proportional to ${\dot n}_\ell(t)$.
According to the detailed calculations given in Ref.~\cite{NSMOY}, we obtain
\begin{widetext}
\begin{align} \label{transporteq}
	\Dot{n}_\ell(t) = 4g^2 \Real\int_{-\infty}^t\!\!\!dt_1 \sum_{\ell_1\ell_2}\Bigl[
	&\phantom{+\;\,}\bigl\{ n_{\ell_1}n_{\ell_2}(1+n_{\ell}) - (1+n_{\ell_1})(1+n_{\ell_2})n_{\ell} \bigr\}_{t_1}
	\bigl( \by_\ell, \bchi_{yy} \bigr)_t \bigl( \bchi_{yy}, \by_\ell \bigr)_{t_1} \notag\\[-10pt]
	&+\bigl\{ n_{\ell_1}(1+n_{\ell_2})(1+n_{\ell}) - (1+n_{\ell_1})n_{\ell_2}n_{\ell} \bigr\}_{t_1}
	\bigl( \by_\ell, \bchi_{yz} \bigr)_t \bigl( \bchi_{yz}, \by_\ell \bigr)_{t_1} \notag\\
	&+\bigl\{ (1+n_{\ell_1})n_{\ell_2}(1+n_{\ell}) - n_{\ell_1}(1+n_{\ell_2})n_{\ell} \bigr\}_{t_1}
	\bigl( \by_\ell, \bchi_{zy} \bigr)_t \bigl( \bchi_{zy}, \by_\ell \bigr)_{t_1} \notag\\
	&+\bigl\{ (1+n_{\ell_1})(1+n_{\ell_2})(1+n_{\ell}) - n_{\ell_1}n_{\ell_2}n_{\ell} \bigr\}_{t_1}
	\bigl( \by_\ell, \bchi_{zz} \bigr)_t \bigl( \bchi_{zz}, \by_\ell \bigr)_{t_1} 
	\Bigr]\,,
\end{align}
\end{widetext}
where 
\begin{align}
	\chi^\alpha_{yy} &= 
	  \zeta^{\Bar\alpha}y_{\ell_1}^\alpha y_{\ell_2}^\alpha
	+ \zeta^{\alpha}y_{\ell_1}^{\Bar\alpha} y_{\ell_2}^\alpha
	+ \zeta^{\alpha}y_{\ell_1}^\alpha y_{\ell_2}^{\Bar\alpha}
	\,,\\
	\chi^\alpha_{yz} &= 
	  \zeta^{\Bar\alpha}y_{\ell_1}^\alpha z_{\ell_2}^\alpha
	+ \zeta^{\alpha}y_{\ell_1}^{\Bar\alpha} z_{\ell_2}^\alpha
	+ \zeta^{\alpha}y_{\ell_1}^\alpha z_{\ell_2}^{\Bar\alpha}
	\,,\\
	\chi^\alpha_{zy} &= 
	  \zeta^{\Bar\alpha}z_{\ell_1}^\alpha y_{\ell_2}^\alpha
	+ \zeta^{\alpha}z_{\ell_1}^{\Bar\alpha} y_{\ell_2}^\alpha
	+ \zeta^{\alpha}z_{\ell_1}^\alpha y_{\ell_2}^{\Bar\alpha}
	\,,\\
	\chi^\alpha_{zz} &= 
	  \zeta^{\Bar\alpha}z_{\ell_1}^\alpha z_{\ell_2}^\alpha
	+ \zeta^{\alpha}z_{\ell_1}^{\Bar\alpha} z_{\ell_2}^\alpha
	+ \zeta^{\alpha}z_{\ell_1}^\alpha z_{\ell_2}^{\Bar\alpha}
	\,,
\end{align}
with $\zeta^\alpha(x)=\BM \zeta(x) \\ \zeta^*(x) \EM^{\alpha}$, 
and $\Bar\alpha$ denotes $\Bar\alpha = 2, 1$ for $\alpha = 1, 2$, respectively.
The subscripts $t$ and $t_1$ are the time arguments, 
and the parenthesis denotes the inner product:
\be
	\bigl( \by, \bchi \bigr)_t = \INT d^3x\;  y^{*,\alpha}(x) \chi^\alpha(x)\,.
\ee
The term in the fourth line of Eq.~(\ref{transporteq}) is what we call the triple production term.
As is discussed in Ref.~\cite{NSMOY}, 
this term plays a crucial role if there is the Landau instability,
but  is vanishing due to the energy conservation otherwise.
While the transport equations with a phase-space distribution, derived in the other methods
 previously, lack this term, the prediction of our transport equation is different
from those based on the other transport equations, when there is the Landau instability.
 
The condition (ii) is the energy renormalization which determines $\delta T$. 
Since the leading contribution to the self-energy is the tag diagram 
 in Fig.~\ref{fig:propagator}(a) and has the form of 
$S_{\ell\ell'}(t_1,t_2) = S'_{\ell\ell'}(t_1) \delta(t_1-t_2)\,$, 
the on-shell self-energy is naturally defined as $S'_{\ell\ell}(t_1)$ 
even in the nonequilibrium situation.
The concrete forms for the self-energy of Fig.~\ref{fig:propagator}(a) and (c) are
\begin{align}
	S^{(\text{a}), 1111}_{\ell\ell}(t_1,t_2) &= \delta(t_1-t_2) \Bigl( \by_\ell, \sigma_3 T^{(a)} \by_\ell\Bigl)_{t_1} \,,\\
	S^{(\text{c}), 1111}_{\ell\ell}(t_1,t_2) &= \delta(t_1-t_2) \Bigl( \by_\ell, \sigma_3 \delta T \by_\ell\Bigl)_{t_1} \,,
\end{align}
with
\be
	T^{(\text{a}), \alpha\beta} (x) = g 
	\BM 2\bra\phi^\d(x)\phi(x)\ket & \bra\phi(x)\phi(x)\ket \\ 
	-\bra\phi(x)\phi(x)\ket^* & -2\bra\phi^\d(x)\phi(x)\ket\EM^{\alpha\beta}\,.
\ee
Then, we find the condition (ii) is satisfied by
\be
	\delta T(x) = T^{(\text{a})}(x) \,,
\ee
which is equivalent to the result of the Hartree-Fock-Bogoliubov approximation.
The matrix elements of $T^{(\text{a})}$ can be rewritten explicitly in terms of $n_\ell(t)$ and $\by_\ell(x)$ as
\begin{align}
	\bra\phi^\d\phi\ket &= \sum_\ell \left[ n_\ell\left(|y_\ell^1|^2 + |y_\ell^2|^2\right) + |y_\ell^2|^2 \right] \,,\\
	\bra\phi\phi\ket    &= \sum_\ell (2n_\ell+1) y_\ell^1 y_\ell^{2,*}\,.
\end{align}
One can show from Eq.~(\ref{TDBdGfory}) that the time-dependence of the total 
number of non-condensed atoms $N_{\mathrm{ex}}(t)=\int d^3x \bra\phi^\d(x)\phi(x)\ket$ is
\be \label{d/dt_Nex}
	\frac{d}{dt} N_{\mathrm{ex}}(t) =  \sum_\ell\left[(2n_\ell(t)+1)\Imag \bigl( \by_\ell, T_0\by_\ell \bigr)
+\Dot{n}_\ell(t) \bigl( \by_\ell, \by_\ell \bigr) \right]\,,
\ee
where $\delta T(x)$ drops because of $\Imag \bigl( \by_\ell, \delta T\by_\ell \bigr)=0$. 
Note that the second term in Eq.~(\ref{d/dt_Nex}) is of two-loop order, 
although we collect only the one-loop diagrams.
This comes from the fact that $\Dot{n}_\ell$ is of one-loop order according to the quantum transport equation (\ref{transporteq}), while $n_\ell$ itself is of no loop one.

\begin{figure}[tb]
\vspace{0.5cm}
\includegraphics[width=8cm]{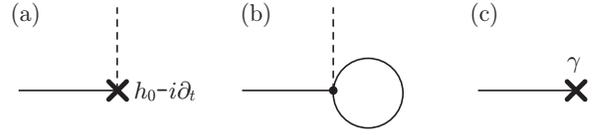}
\caption{\footnotesize{%
Tadpole diagrams.
}}
\label{fig:TagDiagram}
\end{figure}
The last condition (iii) is the self-consistent criterion for dividing the original field operator $\psi$ 
into $\zeta$ and $\phi$. 
The corresponding diagrams at tree and one-loop levels are shown in Fig.~\ref{fig:TagDiagram} (a) and (b), respectively.
Although the energy renormalization is performed at one-loop level, 
the two-loop order correction indicated in Fig.~\ref{fig:TagDiagram} (c) is 
also considered here for the conservations of the total number of atoms. 
The quantity $\gamma(x)$ which is a two-loop order part of the counter term $\delta C(x)$ 
is determined later to cancel the two-loop contribution which appears in Eq.~(\ref{d/dt_Nex}).
The condition with the diagrams in Fig.~\ref{fig:TagDiagram} is written as
\begin{widetext}
\begin{align}\label{tadpole}
	&\int\!\!d^4x_1\; \left( \Delta^{\mu\alpha11}(x,x_1) + \Delta^{\mu\alpha21}(x,x_1) \right) \notag\\
	&\hspace{1cm}\times\biggl[ \Bigl\{h_0(x_1) -i\partial_{t_1} + 2g\bra\phi^\d(x_1)\phi(x_1)\ket\Bigr\}\zeta(x_1) 
	+ g\bra\phi(x_1)\phi(x_1)\ket\zeta^*(x_1) -i\gamma(x_1)\biggr] \notag\\
	-&\int\!\!d^4x_1\; \left( \Delta^{\mu\alpha12}(x,x_1) + \Delta^{\mu\alpha22}(x,x_1) \right)\notag\\
	&\hspace{1cm}\times\biggl[ \Bigl\{h_0(x_1) +i\partial_{t_1} + 2g\bra\phi^\d(x_1)\phi(x_1)\ket\Bigr\}\zeta^*(x_1) 
	+ g\bra\phi(x_1)\phi(x_1)\ket^*\zeta(x_1) +i\gamma(x_1)\Bigr] = 0 \,,
\end{align}
\end{widetext}
which implies
\be \label{TDGP}
	i\tpartial \zeta = \bigl( h_0 + 2g\bra\phi^\d\phi\ket \bigr)\zeta
	+ g\bra\phi\phi\ket\zeta^* - i\gamma\,,
\ee
where $h_0$ is defined in Eq.~(\ref{eq:defh0}).
Thus, the modified TDGP equation has been derived.
To find the function
 $\gamma(x)$, we employ the $\Phi$ derivative approximation \cite{Baym, Kita}, 
which can derive the conserving TDGP equation.
In this approximation, the self-energy is redefined by the derivative of the functional $\Phi = \Phi[G]$ as
\be
	\Sigma(x_1,x_2) = \frac{\delta \Phi}{\delta G(x_1,x_2)}\,.
\ee
Because the two-loop order modification caused by $\dot{n}_\ell$, 
or the thermal counter term $\Hat{Q}$ in other words, is required here, 
we only calculate the part of $\Phi$ which is related to $\Hat{Q}$ in the leading order.
The contribution of the thermal counter term to the self-energy is
\begin{multline}
	\Sigma_Q^{\mu\alpha\nu\beta}(x_1, x_2) = 
	\sum_{\ell_1,\ell_2}\Bigl[ \mathcal{W}_{\ell_1}^{-1}(x_1) \mathcal{B}_{\ell_1}^{-1}(t_1)  \\\times
	S_{Q, \ell_1 \ell_2}(t_1,t_2) \mathcal{B}_{\ell_2}(t_2) \mathcal{W}_{\ell_2}(x_2) \Bigr]^{\mu\alpha\nu\beta}
\end{multline}
with
\be
	S_{Q,\ell_1\ell_2}^{\mu\alpha\nu\beta}(t_1,t_2) =
	-i\Dot{n}_{\ell_1}(t_1)\delta(t_1-t_2) \delta_{\ell_1\ell_2} 
	\left(\begin{array}{cc|cc}
		 0&1& & \\ 0&0 & & \\ \hline & & 0 & 0 \\ & &1 &0 
	\end{array}\right)^{\mu\alpha\nu\beta} \!,
\ee
and that to the functional $\Phi$ is
\begin{align}
	\Phi_Q &= \INT d^4x_1 d^4x_2\; \Sigma_Q^{\mu\alpha\nu\beta}(x_1,x_2) \Delta^{\nu\beta\mu\alpha}(x_2, x_1) \ve^\mu \ve^\alpha \\
	&= -\sum_\ell \INT dt \; \Dot{n}_\ell(t) \bigl( \by_\ell, \by_\ell \bigr)_t \,,
\end{align}
with the sign factor, $\ve^1 =1$ and $\ve^2=-1$.
According to the $\Phi$ derivative approximation, $\gamma(x)$ which is the two-loop 
correction involved 
in $\Sigma_Q$ is found to be
\be
	\gamma(x) = \frac{\delta \Phi_Q}{\delta \zeta^*(x)} 
	= -\sum_\ell\left[ \frac{\delta\Dot{n}_\ell(t)}{\delta\zeta^*(x)} \; \bigl( \by_\ell, \by_\ell \bigr)_t \right]\,.
\ee
By substituting Eq.~(\ref{transporteq}) into this, we obtain
\begin{widetext}
\begin{align}
	\gamma(x) &= g^2 \Real \int_{-\infty}^t \!dt_1\; \sum_{\ell_1 \ell_2 \ell_3} 
\bigl( \by_{\ell_3}, \by_{\ell_3} \bigr)_t \;\biggl[\notag\\
& \left\{ n_1 n_2 (1+n_3) - (1+n_1)(1+n_2)n_3 \right\}_{t_1} X_{yyz}^1(x) \bigl( \bchi_{yy}, \by_3 \bigr)_{t_1}\notag\\
&+\left\{ n_1 (1+n_2) (1+n_3) - (1+n_1)n_2n_3 \right\}_{t_1} X_{yzz}^1(x) \bigl( \bchi_{yz}, \by_3 \bigr)_{t_1}\notag\\
&+\left\{ (1+n_1) n_2 (1+n_3) - n_1(1+n_2)n_3 \right\}_{t_1} X_{zyz}^1(x) \bigl( \bchi_{zy}, \by_3 \bigr)_{t_1}\notag\\
&+\left\{ (1+n_1)(1+n_2)(1+n_3) - n_1n_2n_3 \right\}_{t_1} X_{zzz}^1(x) \bigl( \bchi_{zz}, \by_3 \bigr)_{t_1}
\biggr]\,,
\end{align}
\end{widetext}
where
\be
	X_{yyz}^\alpha = 
	y_1^\alpha y_2^\alpha z_3^{\Bar\alpha} +
	y_1^\alpha y_2^{\Bar\alpha} z_3^\alpha +
	y_1^{\Bar\alpha} y_2^\alpha z_3^\alpha \,,
\ee
and in similar fashions for $X_{yzz}^\alpha$, $X_{zyz}^\alpha$, and $X_{zzz}^\alpha$.
Then, the time derivative of the total number of condensed atoms $N_0(t) = \INT d^3x\; |\zeta(x)|^2$ becomes
\be
	\frac{d}{dt} N_0 = -\sum_\ell\left[(2n_\ell+1)\Imag \bigl( \by_\ell, T_0\by_\ell \bigr)
+\Dot{n}_\ell \bigl( \by_\ell, \by_\ell \bigr) \right]\,,
\ee
which cancels Eq.~(\ref{d/dt_Nex}) and implies the conservation of the total atom number.

Thus, we obtain the coupled equations which describe the temporal evolution of the condensed system, those are
the TDGP equation (\ref{TDGP}), 
the TDBdG equations (\ref{TDBdGfory}), 
and the quantum transport equation (\ref{transporteq}).

\section{Numerical Result}
In this section, we calculate the coupled equations numerically 
to illustrate the nonequilibrium dynamics, 
especially the condensate decays with either the Landau instability or the dynamical one, 
and confirm the qualitative difference between both the instabilities.
For this propose, we consider a very simple system with the one-dimensional Bose--Hubbard model
with the Hamiltonian
\be
	H = \sum_i \left[ -J \psi^\d_i \left\{ \psi_{i+1}+\psi_{i-1} \right\} - 
	\mu \psi^\d_i\psi_i + \frac{U}{2} \psi^\d_i\psi^\d_i\psi_i\psi_i \right]\,.
\ee
Here $J$, $U$, and $i$ represent the inter-site hopping, the on-site couping, and the site index, 
and we put the number of sites $I_s=21$, the total number of atoms $N = 210$, and $U/J = 0.05$.
The condensate is introduced as
$
	\psi_i = \zeta_i + \phi_i
$
with the criterion 
$
	\bra \phi_i \ket = 0\,.
$
It is straightforward to apply the method developed in the previous section 
to this model and to derive the coupled equations for this system in the leading order.

To illustrate the condensate decay with the Landau instability or the dynamical one, 
we consider the following situation. 
First, the equilibrium state with no condensate flow at a temperature $T_0$ is prepared.
Then, the condensate is forced to flow instantaneously with the quasimomentum $k$ : $\zeta_i \to \zeta_i e^{ikx_i}$, 
and the system turns into nonequilibrium.
This nonequilibrium state is chosen as the initial state of the calculation, 
and then the coupled equations are calculated numerically.

\begin{figure}[thb]
\center
\includegraphics[width=4cm]{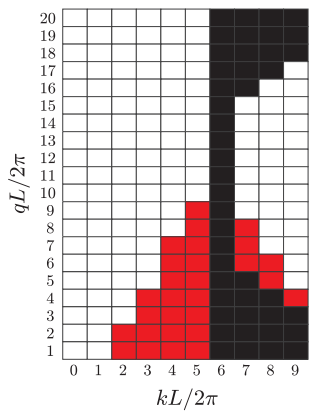}
\caption{\footnotesize{%
Stability diagram for $U/J = 0.05$ obtained analytically from BdG 
eigenvalue equation at tree-level. The symbols $k$ and $q$ stand for the quasi-momenta of the
 condensate and the excitation, respectively, 
and $L$ is the length of the system.
The white, red, and black colored cells denote stable, Landau unstable, and dynamically unstable 
regions, respectively.
}}
\label{fig:PhaseDiagram}
\end{figure}

Solving the BdG eigenvalue equations at zero temperature analytically, 
we obtain the stability diagram as Fig.~\ref{fig:PhaseDiagram} and 
find that the system is stable for $kL/2\pi = 0$ and $1$, Landau 
unstable for $kL/2\pi = 2$ to $5$, 
and dynamically unstable for $kL/2\pi = 6$ to $9$.
Although the result is for the zero temperature, 
it is also expected to be valid for the nonequilibrium case with a sufficiently small initial depletion.

\begin{figure}[tb]
\center
\includegraphics[width=8cm]{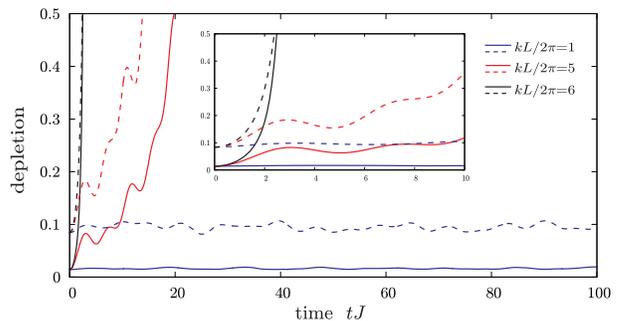}
\caption{\footnotesize{%
Temporal evolution of the depletion $\frac{1}{N} \sum_i \bra \phi_i^\d \phi_i\ket$.
Solid (dashed) line indicates the result with smaller (larger) initial depletion, 
corresponding to a lower (higher) $T_0$.
Blue, red, and black lines correspond to the stable, the Landau unstable, 
and the dynamically unstable conditions, respectively.
}}
\label{fig:depletion1}
\end{figure}
\begin{figure}[tb]
\center
\includegraphics[width=5cm]{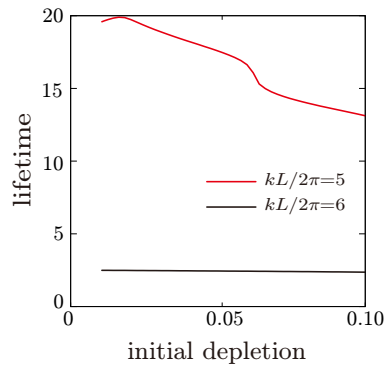}
\caption{\footnotesize{%
Lifetime of the condensate vs initial depletion.
Red and black lines correspond to Landau and dynamically unstable conditions, respectively.
The lifetime is simply defined by the time when the depletion reaches to $0.5$. 
The nonmonotonicity of the lifetime in case of the Landau instability
is due to the oscillating growth of the depletion.
}}
\label{fig:lifetime}
\end{figure}

\begin{figure}
\center
\includegraphics[width=8cm]{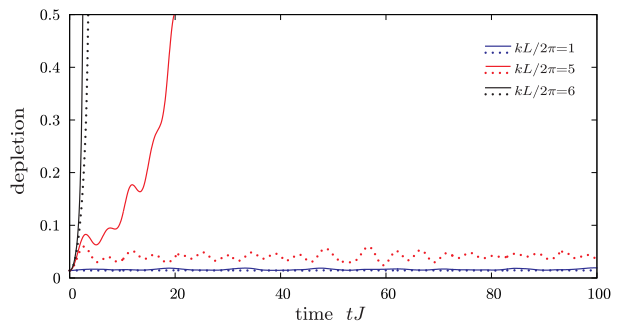}
\caption{\footnotesize{%
The effect of the triple production term to the depletion growth.
Solid line is the same as in Fig.~\ref{fig:depletion1}, 
while doted line indicates the result, obtained when the triple production term is omitted.
}}
\label{fig:depletion2}
\end{figure}
The coupled equations are calculated numerically with several initial conditions. 
The depletion of the condensate is indicated in Fig.~\ref{fig:depletion1}, and
the stable and unstable behaviors are clearly discriminated.
For the stable condition $kL/2\pi=1$, the nonequilibrium depletion
oscillates around the equilibrium one and never grows.
The depletion grows initially with oscillation for the Landau unstable condition $kL/2\pi=5$,
but grows much more rapidly and 
exponentially from the beginning for the dynamically unstable one $kL/2\pi=6$.
The decay speed with the Landau instability tends to increase 
if the initial depletion becomes large 
as is shown in Fig.~\ref{fig:lifetime}, reflecting the fact that
 the Landau instability is caused by 
the collision between condensate and non-condensate particles and that the collision
becomes more frequent for larger depletion.
On the other hand, the value of the initial depletion is not relevant for the dynamical
 instability, since the non-condensate particle plays no essential role then. 

The effect of the triple production term which is a distinguishing feature of
 our quantum transport equation is 
shown in Fig.~\ref{fig:depletion2}.
No qualitative difference between the cases with the triple production term and without it
is notable in both the stable and dynamically unstable conditions.
That is because that the triple production is suppressed due to the energy conservation
 in the former,
and that the rapidly growth is governed by the TDGP and TDBdG equations
 but not by the transport equation in the latter.
In the Landau instability condition, on the other hand, the difference is remarkable.
Basically, the growth behavior disappears if the triple production term is omitted.
Thus we conclude the essential contribution of the triple production term
 in describing the Landau instability.

Although we present only the results of the three typical values of $k$, representing 
the stable, Landau unstable, and dynamically unstable situations,  respectively,
the qualitatively similar results are obtained for any allowed value of $k$.

\section{Summary}
In this paper, the self-consistent equations which describe the nonequilibrium dynamics 
have been derived by applying the nonequilibrium TFD to the condensed atom system.
The system with a time-dependent order parameter is considered as 
the further extension of our previous study where the stationary order parameter has been assumed.
We only have been able to describe the first stage of the condensate decay with the Landau instability heretofore.
Now we can predict the next stage of the decay dynamics, 
and can describe not only the Landau instability but also the dynamical one.

To treat the time-dependence of the order parameter within the time-independent quasiparticle picture, 
the field operator is expanded with the complete set evaluating by the TDBdG equations.
This method, proposed first by Matsumoto and Sakamoto without a detailed vindication, 
is obtained here to construct the particle operator time-independent.
The renormalization conditions, the the Chu-Umezawa's diagonalization condition (i), the energy renormalization (ii), 
and the criterion for diving the non-condensate and the condensate (iii), 
are applied to determine the coupled equations which are the quantum transport equation, 
the TDBdG equations, and the TDGP equation.

The point is that the coupled equations we obtained are not the naive assemblages of presumable equations, 
but are the self-consistent ones derived by the appropriate renormalization conditions.
To confirm that our coupled equations can describe both Landau and dynamical instability, 
we consider an one-dimensional Bose--Hubbard model, 
and calculate numerically the depletion whose growth implies the decay of the condensate.
Since the initial depletion is sufficiently small, 
the unstable condition is well characterized by the eigenvalues of the BdG equations at tree-level.
We found that the depletion growths with both the Landau and dynamically unstable condition, while dose not with stable condition.
Predictably in the Landau unstable condition, the growth is slower and has stronger initial depletion dependence 
than that in the dynamically unstable condition.

As we reported previously, our transport equation contains the triple production term 
which is absent in the one of the other methods.
The difference originates in the choice of the quasiparticle picture which is essential for the quantum field theory.
We construct the quasiparticle faithfully to the quantum field theory, while
the others are based on a phase-space distribution function and no explicit particle representation is given.
This difference is inconspicuous if there is no Landau instability, 
but causes a great qualitative change if there is Landau instability. 
Basically, we find that the condensate dose not decay without 
the triple production term in the Landau unstable condition, 
although the decay has been observed experimentally. 
We emphasize that the choice of the quasiparticle picture is even more essential for a unstable case.
A quantitative comparison to a experiment of the condensate decay will be the future task.

\begin{acknowledgments}
The authors thank the Yukawa Institute for Theoretical Physics at Kyoto University 
for offering us the opportunity to discuss this work during the YITP workshop 
YITP-W-09-12 on ``Thermal Quantum Field Theories and Their Applications''.
\end{acknowledgments}

\end{document}